\documentclass{article}
\input epsf.sty
\newcommand{\bright}{\begin{flushright}}
\newcommand{\eright}{\end{flushright}}
\newcommand{\bminip}{\begin{minipage}}
\newcommand{\eminip}{\end{minipage}}
\newcommand{\bcent}{\begin{center}}
\newcommand{\ecent}{\end{center}}

\newcommand{\nnb}{\nonumber}
\newcommand{\reflef}{(\ref}
\newcommand{\MP}{M_{\rm P}}
\newcommand{\lmd}{\lambda}
\newcommand{\Lmd}{\Lambda}
\newcommand{\gsim}{\mbox{\raisebox{-.3em}{$\;\stackrel{>}{\sim}\;$}}}
\newcommand{\lsim}{\mbox{\raisebox{-.3em}{$\;\stackrel{<}{\sim}\;$}}}

\newcommand{\beq}{\begin{equation}}
\newcommand{\eeq}{\end{equation}}
\newcommand{\bfig}{\begin{figure}}
\newcommand{\efig}{\end{figure}}

\begin{document}

\baselineskip=0.6cm
\mbox{}\\[-3.5em]

\begin{center}
{\Large\bf How strongly does dating meteorites constrain the time-dependence of the fine-structure constant? 
}\\[.6em]
Yasunori Fujii$^1$ and Akira Iwamoto$^2$\\[.6em]
{\small
\hspace*{-.4em}$^1$Advanced \hspace{-.2em}Research \hspace{-.2em}Institute \hspace{-.2em}for \hspace{-.2em}Science \hspace{-.2em}and \hspace{-.2em}Engineering, \hspace{-.2em}Waseda \hspace{-.2em}University, \hspace{-.2em}Shinjuku, \hspace{-.2em}Tokyo, \hspace{-.2em}169-8555 \hspace{-.2em}Japan\\
$^2$Japan Atomic Energy Research Institute (JAERI), Tokai-mura, Naka-gun, Ibaraki, 319-1195 Japan}\\[1.5em]
{\large\bf Abstract}
\end{center}
\mbox{}\\[-.1em]
\baselineskip=.6cm
\begin{center}
\mbox{}\\[-5.8em]
\begin{minipage}{14.1cm}
We review our argument on the nature of the so-called meteorite constraint on the possible time-dependence of the fine-structure constant, emphasizing that dating meteorites at the present time is different in principle from searching directly for the traces in the past, as in the Oklo phenomenon and the QSO absorption lines.  In the related literature, we still find some arguments not necessarily consistent with this difference to be taken properly into account.  It does not immediately follow that any model-dependent approaches are useless in practice, though we cannot help suspecting that dating meteorites is no match for the Oklo and the QSO in probing the time-variability of the fine-structure constant, at this moment.  Some of the relevance to the QSO data particularly in terms of the scalar field will be discussed.

\end{minipage}
\end{center}
\mbox{}\\[-3.5em]

\section{Introduction}

Ever since Dirac's suggestion \cite{dir} that the gravitational ``constant" $G$ may not be truly constant, many measurements as well as theoretical analyses have been made on what had been accepted to be fundamental constants, as the experimental technique has steadily improved in the accuracy.  Unfortunately his prediction $G(t) \sim t^{-1}$ is already excluded by the recent observational findings, $\dot{G}/G = (0.2\pm 0.4)\times 10^{-11}{\rm y}^{-1}$ and $(-0.06 \pm 0.2)\times 10^{-11}{\rm y}^{-1}$ derived from the Viking Project \cite{vik} and the analysis of the binary pulsars \cite{binaryp}, respectively.  They are less than the predicted absolute value $|-t_0^{-1}|\sim 0.7\times 10^{-10}{\rm y}^{-1}$, by at least an order of magnitude, where $t_0\approx 1.37\times 10^{10}{\rm y}$ is the age of the Universe.  An order-of-magnitude difference might look rather trivial compared with a huge number like $t_0$. To appreciate how seriously it might affect the whole scenario, however, let us try to understand the above results by assuming an approximate equation $\dot{G}/G =-0.1 \times t_0^{-1}$, indicating $G(t)\sim t^{-0.1}$, which varies too slowly compared with the behavior $\sim t^{-1}$, as will be shown.

Dirac started with noticing a near coincidence between the two large numbers, today's value of the ratio of the electromagnetic to the gravitational coupling strengths of the electrons ${\cal R}(t_0)\equiv \alpha/(Gm_{\rm e}^2)=8\pi \alpha (\MP /m_{\rm e})^2 \approx 4.3\times 10^{42}$ and $t_0/t_{\rm e}\approx 3.3\times 10^{38}$, where $\alpha=e^2/(4\pi \hbar c)$ is the fine-structure constant, while $\MP = (8\pi G/(c\hbar))^{-1/2}$ and $m_{\rm e}$ are the Planck mass and the electron mass, respectively. Also $t_{\rm e}=\hbar/(m_{\rm e}c^2)= 1.29\times 10^{-21}{\rm sec}$ is one of the fundamental microscopic times.  By assuming $G(t)\sim t^{-1}$, he readily obtained a reasonably small value for ${\cal R}(t_{\rm e})\approx 1.3\times 10^4$, sufficiently close to order unity, theoretically a more natural number than ${\cal R}(t_0)$.  His ``success" will be lost, however, if $G(t)\sim t^{-0.1}$, yielding ${\cal R}(t_{\rm e})\sim 0.6\times 10^{39}$,  far from being anything close to one.

These ``failures" have, however, never deterred physicists from pursuing the idea of time-dependent constants.  For example, ${\cal R}(t_0)$, one of the large numbers discussed above is no longer a serious issue because we are now familiar with the renormalization-group equations expressing the ``hierarchy ratio" in terms of an exponential function, $\MP/m_{\rm e} \approx \exp (b/(2\alpha))$ with a comfortable choice $b\sim 0.74$.

On the other hand, the current idea of ``unification" has its own reason why we have to be concerned about possible time-dependent coupling constants among fields.  In the fundamental theory in higher dimensions we may start with assuming truly constant coupling constants, which descend to the observed coupling constants in the real 4-dimensional world on being multiplied by a scalar field, like the one representing the size of internal space in the Kaluza-Klein theories \cite{KK,sahdev}, or the ``dilaton" in string theory \cite{string,callan}.  Some of them might be identified as dark energy supposed to be responsible for the acceleration of the Universe \cite{cup}--\cite{beltor}.   We expect that such a field evolves slowly with the cosmic time, and so do the observed coupling constants, in a way not necessarily like $t^{-1}$.  In this context, detecting time-variability of any of the coupling constants might even be viewed as an ``evidence" of the presence of an underlying theory at the deeper level behind the world as we see it.

Among various coupling constants and some of the mass ratios, the suspected time-variability of the fine-structure constant has been a special focus of the recent studies, because it is not only dimensionless but also allows an easier and closer access than others.  Laboratory experiments belong obviously to this class of attempts, though they are designed to look at almost instantaneous rate of change at the present time \cite{labexp}.  Measurements over much longer time-spans include the analysis of the Oklo phenomenon and the observation of the absorption lines from quasi-stellar objects (QSO).

The Oklo phenomenon occurred at $(1.95\pm 0.05) \times 10^9$ years ago \cite{Okloage}, providing an exceptionally stringent constraint \cite{shly}, either an upper bound $\Delta\alpha /\alpha =(-0.8\pm 1.0)\times 10^{-8}$ or a nonzero change $(0.88\pm 0.07) \times 10^{-7}$ \cite{fih}, where $\Delta\alpha$ means the value of $\alpha$ at the Oklo time minus today's value. See the second of Ref. \cite{fih} for the slight revisions of the results in the first paper. The analysis  of the neutron absorption process of $^{149}{\rm Sm}$ exploits an amplification effect due to (today's) very small resonance energy, $E_{r0}=97.3 {\rm meV}$, compared with the typical mass scale of nuclear phenomena of the order of MeV.  Note that the timing corresponds to the fractional look-back time $s=1-t/t_0 \approx 0.142$ with the age of the Universe $t_0 \approx 1.37 \times 10^{10}{\rm y}$, which can also be viewed as the cosmological redshift $z\approx 0.156$, based on the now standard cosmological model with spatially flat 3-space with the cosmological constant parametrized by $\Omega_\Lmd =0.7$ and the Hubble parameter $H_0=72 \:{\rm km/sec/Mpc}$ \cite{WMAP}.

Even with possible complications due to the strong interaction \cite{fritzsch,olive1}, it is quite unlikely that the upper bound  turns out to be an order of magnitude larger than the Coulomb-only estimate, as was discussed in Appendix A of \cite{yfptb}, in which we only assumed the nearly complete cancellation between the energies from the Coulomb and the strong interactions for any value of $E_r$ unless a very special relation happens to be present between them.  Hence more likely is the value below the Coulomb-only estimate.  The conclusion is sufficiently general to survive including any specific mechanisms, like the time-dependent strange quark mass \cite{flam}, for example, in the strong-interaction contribution.

It has been argued recently that the neutron flux might likely have been in the thermally non-equilibrium state, thus yielding a nonzero result $\Delta\alpha /\alpha =(0.41 \raisebox{1.1ex}{\footnotesize \hspace{.3em}+0.14}\raisebox{-.6ex}{\hspace{-.8em {\footnotesize --0.06}}})\hspace{0em}\times 10^{-7}$ at the Oklo time \cite{nonmax}.  A more detailed analysis has been made basically in the same context \cite{petrov}, reaching the upper bound $(-0.05\pm 0.61)\times 10^{-7}$.  As we find, the computed absorption cross section of samarium-149 falls off toward the negatively large value of $\Delta E_r=E_r- E_{r0}$ for 2 billion years ago, not only for the Maxwellian distribution in Fig. 1 of \cite{fih} but also for the supposedly more realistic neutron spectrum in Fig. 3 of \cite{nonmax} as well as in Fig. 12 of \cite{petrov}.  This makes it likely that the measured and the calculated cross sections meet together for $\Delta E_r \gsim -100\:{\rm meV}$, hence $\Delta\alpha/\alpha \lsim 0.9\times 10^{-7}$, as in our non-null result \cite{fih}.  (No significant departure from the Maxwellian spectrum occurs for positive $\Delta E_r $.)  Combining this with our argument above \cite{yfptb}, we find it unlikely that the Oklo constraint is large enough to be comparable with the current QSO result, as will be shown shortly, even if uncertainties are yet to be taken into account on the details of Reactor Zones (RZ) 10 and 13, from which we collected the samples to minimize the possible contamination from outside after the end of the reactor activity \cite{fih}.

\bfig[bft]
\vspace{-3em}
\hspace*{8.6em}
\epsfxsize=9cm
\epsffile{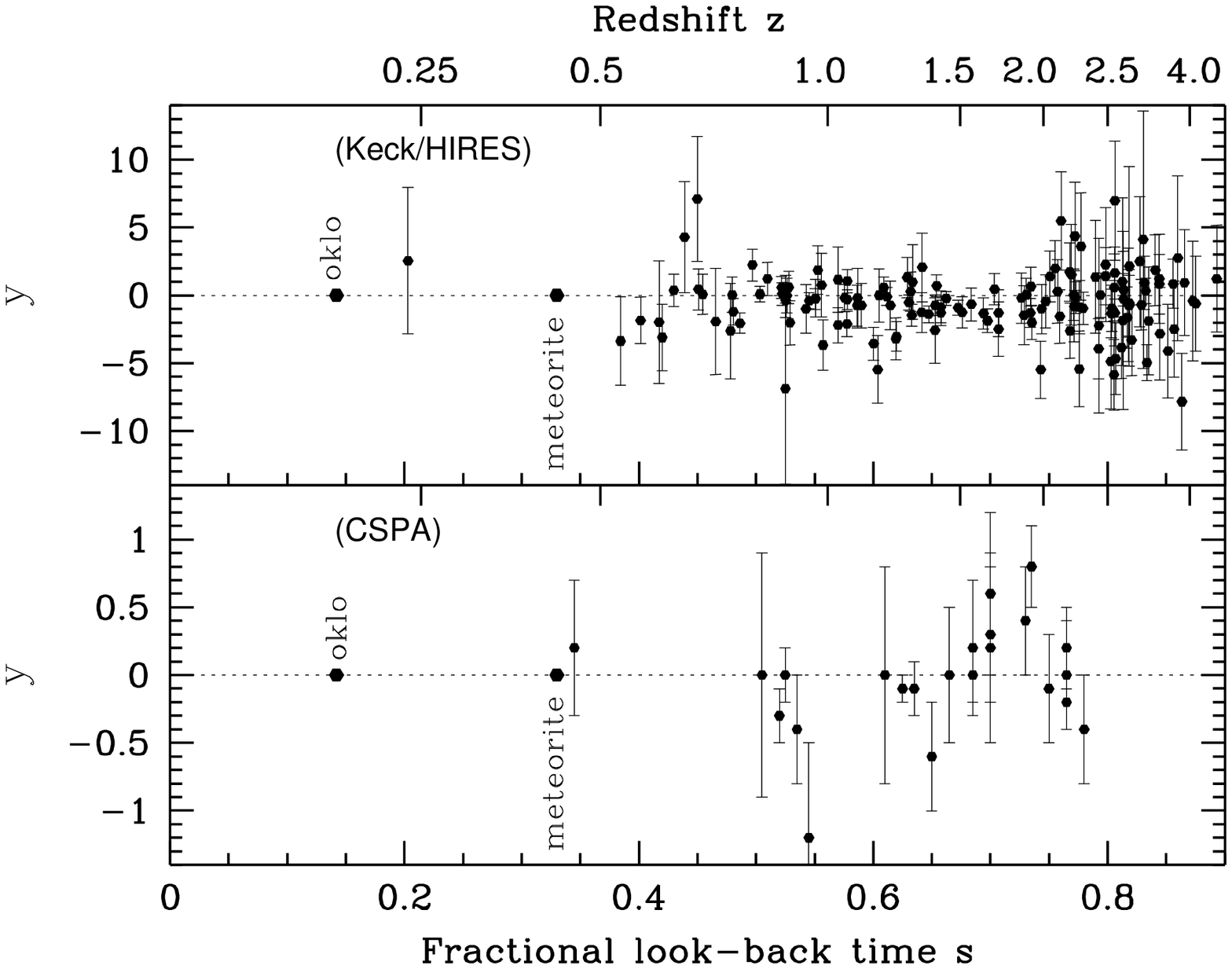}
\mbox{}\\[-8em]
\caption{The upper panel shows $y=(\Delta\alpha/\alpha) \times 10^5$ versus either the fractional look-back time $s=1-t/t_0$ or the redshift $z$ obtained from Ref. \cite{keck} (143 data-points), while the lower panel from Ref. \cite{indfr} (case 1 in the first of the quoted references, 23 data-points). Notice an order-of-magnitude difference in the vertical scales shown on the left.  The 2 data-points from Refs. \cite{qrl,lev} not shown here share nearly the same features as in the lower panel.  Also shown are the Oklo and the so-called meteorite constraints.  The error-bars on the former are invisibly small both in the vertical and horizontal scales, whereas  those for the latter are controversial, being the main subject of this study.}
\label{summ}
\efig

On the other hand, the QSO observations with the technique of the many-multiplet method appear to indicate values at the level of $|\Delta\alpha /\alpha |$ somewhere around $10^{-6}-10^{-5}$, in the ranges $0.20\lsim s \lsim 0.89$, or $0.23 \lsim z \lsim 4.7$ \cite{keck}--\cite{lev}.  See Refs. \cite{chiba,uzan} for reviewing the earlier attempts. The circumstance we are going to deal with may be illustrated in Fig. \ref{summ}, in which we show the two sets of data.  The 143 data-points from Keck/HIRES \cite{keck} provides $\Delta\alpha/\alpha =(-0.573\pm 0.113) \times 10^{-5}$ thus claiming a negatively nonzero change, while the other of 23 data-points from VLT-UVES  \cite{indfr} (case 1 in the first of the quoted references) reveals no such variation with $(-0.06\pm 0.06)\times 10^{-5}$.  Another group \cite{qrl,lev} also from VLT-UVES favors a null result from their much more careful analysis of 2 data-points.  It is even not clear if the contradictions come from some of the systematic uncertainties, or from the lack of sensitivity of the measurement implying that the values observed so far do not represent the true effect supposed to be hidden at a deeper level. For the clarification of the last point of view, building larger telescopes is among future projects in this research area \cite{lev}.

One of the common key issues is an apparent difference between the Oklo constraint and the reported QSO data by (far) more than an order of magnitude, corroborated particularly by our discussion outlined above.  This may have something to do with the fact that the Oklo event lies outside the QSO range, as shown in Fig. \ref{summ}.  By exploiting this situation we may expect that $\Delta \alpha/\alpha$ is not a monotonic function of time, like the  constant values chosen by the simple weighted-mean estimate attempted in the conventional analyses \cite{keck}--\cite{lev}.  It can even be oscillating in such a way that the Oklo time corresponds to one of zeros of the oscillation, also supported by an oscillatory behavior of the scalar field supposed to be responsible for the accelerating Universe \cite{cup}--\cite{plb2}, as will be discussed briefly in the final Section.  In this context, we could wish if we find another kind of phenomena which turn out to be useful to probe time-dependence of $\alpha$.

This is a reason why Dyson's analysis attempted decades ago still
attracts attentions.  He discussed a possible constraint obtained by
examining the very long-lived beta decay $^{187}{\rm Re}\rightarrow
^{187}\!{\rm Os}$ in iron meteorites \cite{dyson}.  By assuming a rather extreme circumstance, he obtained $|\Delta\alpha/\alpha|\lsim 10^{-3}\sim 10^{-4}$ around the early time of the solar system.  The bound is less stringent than those for the Oklo and the QSO which we are now interested in.  This constraint, to be called the ``meteorite constraint" in what follows, is nevertheless interesting because the crucial process might have occurred when the iron meteorites were formed about 4.6 billion years ago, also referred conveniently to the ``meteorite time,"  which, as indicated in Fig. \ref{summ}, lies between the Oklo time and most of the times for the QSO absorption lines.

A possible importance of the rhenium decay appeared to be re-discovered when Olive et al. pointed out \cite{olive1} that the decay rate in question has been measured much more accurately than before, $\lmd \approx 1.666 \times 10^{-11}{\rm y}^{-1}$, with the precision of 0.5\% \cite{smol}.  We believe that this approach needs more scrutiny by taking possible time-dependence of the decay rate into account correctly \cite{yfai} through an extension of the isochron analysis.  We come eventually to a conceptually different approach, finding that the analysis tends to depending on details of how the decay rate varies with time, contrary to the model-independence claimed earlier.  We expect corresponding reassessments to be made in the literature on the related subject, including  Refs. \cite{wagg,olive1,uzan,olive2}.

We nevertheless try to apply the above extended analysis to find a way to reduce inherent uncertainties in any past studies.  We present an explicit theoretical model, but obtaining the result not very much promising, in practice.  These discussions appear to point to the view that dating meteorites may not provide a constraint as strong as the Oklo and the QSO, as was also suggested toward the end of III.A.4 in \cite{uzan} without much elaboration.  This is probably a fair conclusion to be reached at this moment, unless an entirely new ingredient is added, as suggested briefly in the final Section.

In Section 2, we begin with reviewing Dyson's classic work on the decay
of rhenium-187.  We then proceed to discuss the traditional isochron analysis in Section 3, based on which Olive et al. claimed a new result nearly as stringent as the Oklo constraint.  This inspired us to extend the isochron technique in Section 4, by allowing the decay rate $\lmd(t)$ to be time-dependent, due to the time-dependent fine-structure constant, leading us to the conclusion that the crucial role is played by the time-averaged decay rate $\bar{\lmd}$.  The need to differentiate these two different kinds of decay rate is strongly emphasized.  In Section 5, we first re-examine Dyson's analysis in the new light of the extended version of the isochron technique.  We then propose a theoretical model which is expected to derive the change of $\bar{\lmd}$ from that of $\lmd (t)$ based on an explicit but somewhat artificial mechanism, by fully exploiting the observational uncertainties still present in today's isochron analysis. The model turns out to select a time-range around the meteorite time, but with the resulting $\Delta\alpha/\alpha$ short of being unique.  In the final Section 6, we discuss the relevance to the Oklo constraint and the analyses of the QSO absorption lines, together with the related remarks on the scalar-tensor theory.

\section{Dyson's argument -- I}

The exceptionally rare decay of rhenium-187 is a consequence of an exceptionally small mass difference of its beta-decay into osmium-187, $Q=2.5\:{\rm keV}$, compared with the estimated Coulomb energy difference $Q_c =-15.8\:{\rm MeV}$ (according to the mass formula used by Dyson \cite{dyson}), most of which happens to be canceled by the energy due to the strong interaction.  In the Coulomb-only estimate, we make use of the relation $Q_c \propto \alpha$ to derive
\beq
\frac{\Delta\alpha}{\alpha}=\frac{\Delta Q}{Q_c}=\frac{Q}{Q_c}\frac{\Delta Q}{Q}\approx -1.6\times 10^{-4}\frac{\Delta Q}{Q},
\label{dys-1}
\eeq
where $\Delta Q = Q(t)-Q(t_0)$, like the same convention for $\Delta\alpha$.

We find
\beq
\lmd \propto Q^p, 
\label{dys-2}
\eeq
with $p\approx 2.84$ according to \cite{dyson} (though $p=3$ is used in \cite{olive1} leaving most of the results practically unaffected).  We then obtain

\beq
\frac{\Delta\lmd}{\lmd}\approx \left( 1+\frac{Q_c}{Q}\frac{\Delta\alpha}{\alpha}  \right)^p -1,
\label{dys-21}
\eeq
where $\lmd$ is chosen to be today's value of the decay rate $\lmd_0 \approx 1.67\times 10^{-11}{\rm y}^{-1}.$

In the linear approximation with
\beq
\frac{\Delta\lmd}{\lmd}\approx p\frac{\Delta Q}{Q},
\label{dys-22}
\eeq
we use \reflef{dys-1}) to find
\beq
\frac{\Delta\alpha}{\alpha}\approx -5.6\times 10^{-5}\frac{\Delta\lmd}{\lmd},
\label{dys-3}
\eeq
or
\beq
\frac{\Delta\lmd}{\lmd} \approx -1.78\times 10^4 \frac{\Delta\alpha}{\alpha}.
\label{dys-3-1}
\eeq
The occurrence of a small (large) number like $10^{-5}$ ($10^4$) on the right-hand sides represents an amplification effect, as was emphasized in the previous Section on the Oklo constraint.

As Dyson noticed in \reflef{dys-21}), if $\Delta\alpha/\alpha$ is negative and sufficiently large, $\Delta\lmd /\lmd$ on the left-hand side can be so large positively that the decay rate at the time nearly comparable with 4.6 billion years ago should have prompted rhenium to  decay too rapidly to leave enough amount of rhenium as  we observe today on the Earth.  Dyson required this possibility to be ruled out.  Note that, without any such change, we have $(t_0-t_1)\lmd_0 \approx 0.077$, hence $e^{-(t_0-t_1)\lmd_0} \approx 0.926$, implying that most of rhenium present at $t_1$, the time of formation of meteorites, is still present  now, where $t_0-t_1 \approx 4.6\times 10^9{\rm y}$.
\bfig
\vspace{-5em}
\hspace*{11em}
\epsfxsize=7cm
\epsffile{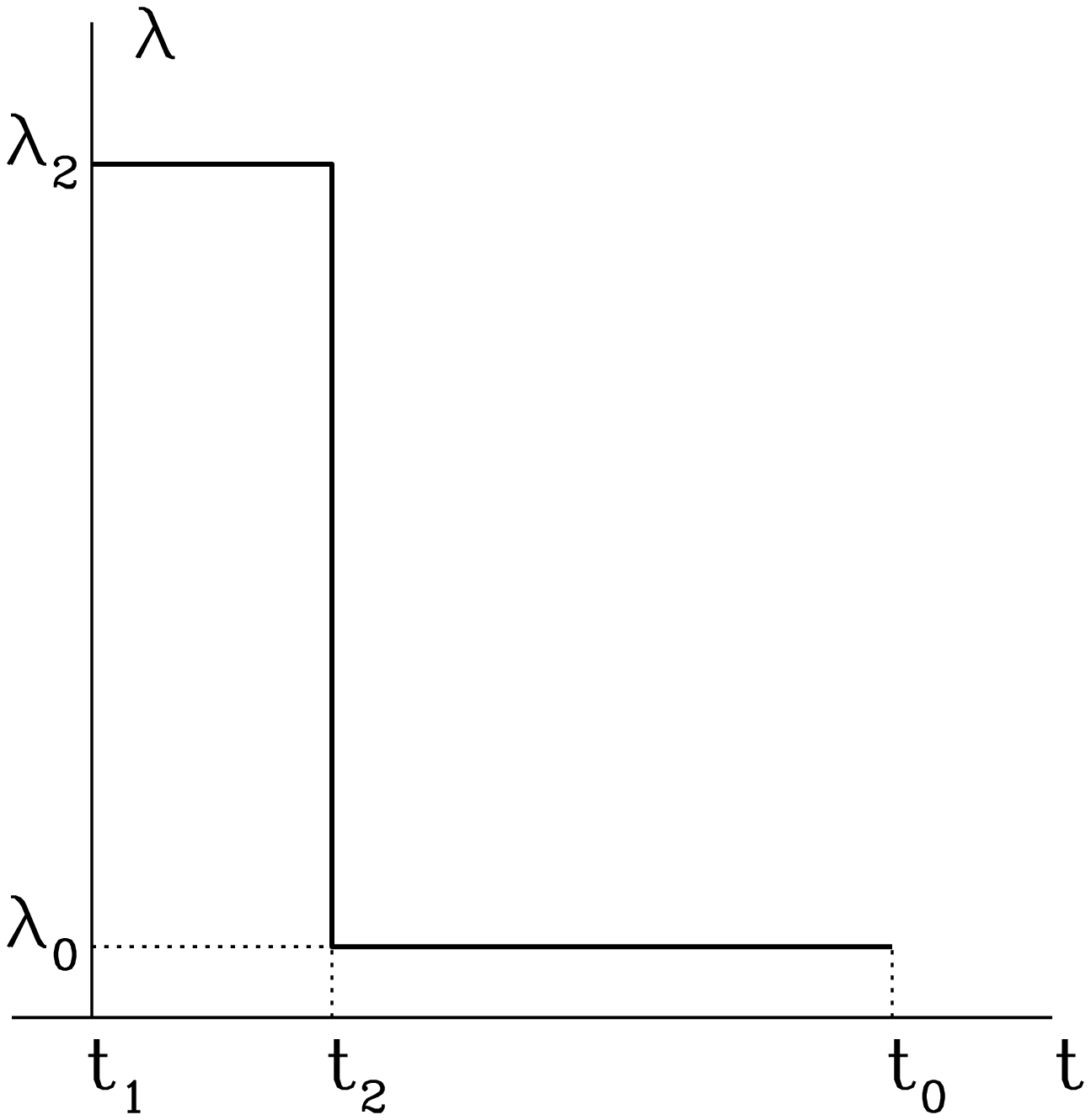}
\mbox{}\\[-5em]
\caption{The decay rate is $\lmd_2$ in the time-interval between $t_1$ and $t_2$.}
\label{tintv1}
\efig

Since $\alpha$ is supposed to be different from today's value only in the past, $\lmd$ can be much larger also in the remote past, probably toward the initial time $t_1$.  As in Fig. \ref{tintv1}, we may simplify the situation by assuming that the decay rate reached a significantly large value $\lmd_2$ only during an interval, $t_1 <t<t_2$, which was the earliest time-interval toward $t_1$:
\beq
\frac{t_2-t_1}{t_0-t_1}= \frac{1}{10},
\label{dys-a3}
\eeq
for example, giving $t_2-t_1 \approx 4.6\times 10^{8}{\rm y}$.  Suppose $\lmd_2 \approx 100 \lmd_0$, then we would find the product as large as $(t_2-t_1)\lmd_2 \approx (100/10) \times 0.077 \approx 0.77$, and hence $e^{-0.77}=0.46$, which is dangerously small to be avoided.  We choose  the above $\lmd_2$ as representing a safe upper bound 
\beq
\frac{\Delta\lmd}{\lmd}\lsim 100.
\label{dys-a3a}
\eeq
Substituting this number into the left-hand side of \reflef{dys-21}), we find 
\beq
-\frac{\Delta\alpha}{\alpha}\lsim 6.5\times 10^{-4}, \quad\mbox{or}\quad\frac{\dot{\alpha}}{\alpha}\lsim 1.4 \times 10^{-13}{\rm y}^{-1}.
 \label{dys-23}
\eeq

This is essentially Dyson's result which we re-interpreted in terms of the time-span $t_1 <t < t_2$, for the sake of later discussion.  In fact we will show that the presence of the new result from dating undermines the approach described above.  Notice also that with the large value \reflef{dys-a3a}) we fail to justify the linear approximation, resulting in $-\Delta\alpha/\alpha \approx 5.6\times 10^{-3}$, which happens to be nearly an order-of-magnitude larger than the value given by \reflef{dys-23}).

\section{Isochron analysis -- traditional}

Meteorites are considered to have been formed nearly at the time when the solar system was created, as fragments of small planets. A single core of each planet was fragmented into a number of different meteorites.  The original planets contained rhenium and osmium, which were created much earlier as supernovae had exploded.  Some of the mother planets of meteorites were melted partially or totally and materials of metal and silicates were separated in the planets.  The portions of liquid metal were crystallized to form solid metal cores, mainly composed of iron thus creating various different iron meteorites.  Different temperatures and pressures, and hence different chemical fractionations provided initial contents of rhenium, for example, which were different from meteorite to meteorite.  Iron meteorites ejected from the same mother planet have similar chemical properties, hence are categorized into the same group. Because the trapped rhenium-187 in iron meteorites decayed slowly into osmium-187, isotope ratios of rhenium and osmium changed gradually with time.

Let $t_1$ denote the time when all the meteorites in the same group were
formed.  In each of them, we define the ratio $N_{\rm Re} = (^{187}{\rm
Re})/(^{188}{\rm Os})$ in terms of the abundance of the stable isotope $^{188}{\rm Os}$.  We find
\beq
N_{\rm Re}(t)=Ae^{-\lmd(t-t_1)},
\label{isch-1}
\eeq
where $A$ for the initial value differs from meteorite to meteorite, while $\lmd$ is for the decay rate.  The decay product reappears as the abundance of $^{187}{\rm Os}$ also defined as the ratio to $^{188}{\rm Os}$:
\beq
N_{\rm Os}(t)=A\left( 1-e^{-\lmd(t-t_1)} \right) + B,
\label{isch-2}
\eeq
where $B$ represents the common amount having been present already at $t=t_1$.  From \reflef{isch-1}) and \reflef{isch-2}) the coefficient $A$ is eliminated giving
\beq
N_{\rm Os}(t)=S(t)N_{\rm Re}(t) +B,
\label{isch-3}
\eeq
where the time-dependent $S(t)$ is defined by
\beq
S(t)=e^{\lmd(t-t_1)}-1,
\label{isch-4}
\eeq
which starts with zero, increasing as $t-t_1$.

The advantage of using the expression \reflef{isch-3}) will be appreciated if we measure the abundances of $^{187}{\rm Re}$ and $^{187}{\rm Os}$ and plotted them in a 2-dimensional plot as in Fig. \ref{figisch1}, often called an isochron plot.  If they are found to align along a single straight line with a common slope $S(t)$ and the intercept $B$, we may justify that the meteorites were born at the same time, hence belonging to the same group.   The obtained value of the slope at the present time determines the product $\lmd (t_0-t_1)$.  If the value of $\lmd$ is known, then we find the age $t_0-t_1$.  This is the principle of dating.

\bfig[htb]
\vspace{-16em}
\epsfxsize=11.cm
\hspace*{7.5em}
\epsffile{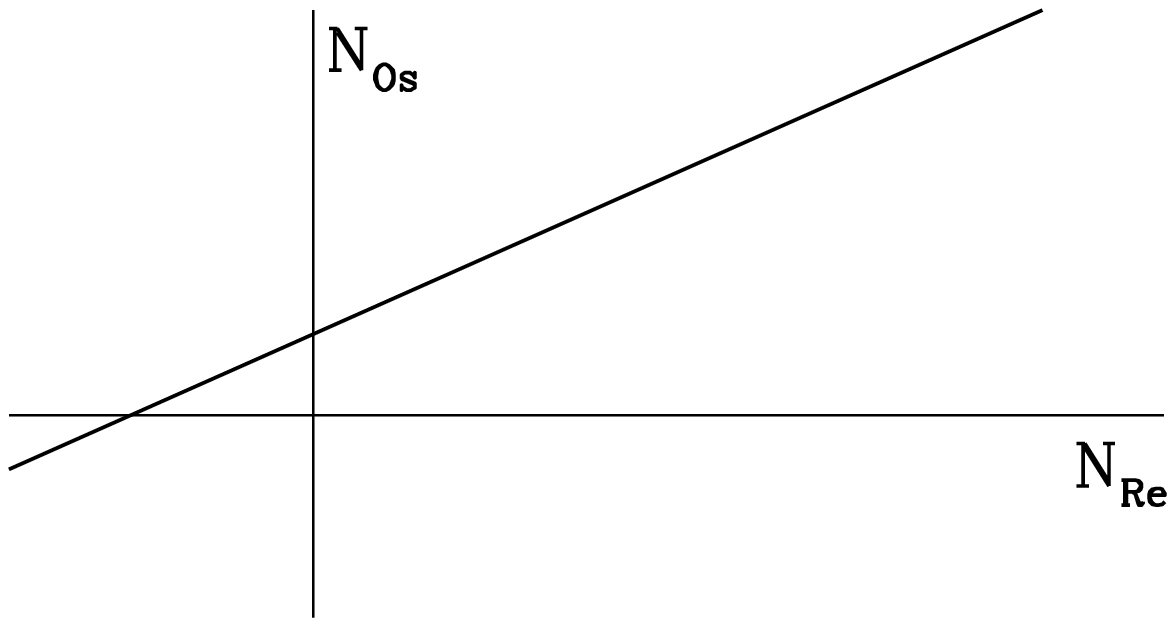}
\mbox{}\\[-5em]
\caption{2-dimensional plot of the Re-Os isochron.}
\label{figisch1}
\efig
\mbox{}\\[-3em]

Smoliar et al. reported the latest development of dating iron meteorites \cite{smol}. They applied a very careful analysis of the data obtained from the four groups (IIA,IIIA,IVA, and IVB), determining the slope with the precision of 0.5\%.  On the other hand, the ages of the angrite meteorites assumed to be formed at about the same time as the iron meteorites have been known with better precision by using the same technique applied to the U-Pb, and Mn-Cr isochron pairs,  partly due to much larger decay rates.  Accepting the absolute age  $t_0-t_1 =(4.5578\pm 0.0004) \times 10^9 {\rm y}$ determined in this way \cite{angrite}, we then find $\lmd$ for the Re-Os decay to be $\lmd \approx 1.666\times 10^{-11}{\rm y}^{-1}$, with the precision of 0.5\%, as mentioned before.  This might be denoted by
\beq
\Bigl| \frac{\Delta\lmd}{\lmd}  \Bigr| \lsim 5\times 10^{-3}.
\label{isch-5}
\eeq

Olive et al. seemed to have expected that this new result implied a new constraint on the time-variability of $\alpha$ \cite{olive1}, by assuming that the left-hand side of \reflef{isch-5}) is the same as, or at least the upper bound of the largest possible time-change of $\lmd(t)$ due to the time-dependent $\alpha(t)$.  They substituted \reflef{isch-5}) into \reflef{dys-3}) to find
\beq
\Bigl| \frac{\Delta\alpha}{\alpha} \Bigr| \lsim 2.5\times 10^{-7},
\label{isch-6}
\eeq
which turns out to be nearly as stringent as the Oklo constraint, though the same authors later admitted \cite{olive2} that their assumption cannot be justified in practice, conceding that \reflef{isch-6}) is ``over-restrictive."

They also assumed that \reflef{isch-6}) represents a time-variation that occurred around $t_1$.  This intrigued us because the time is somewhat older than the Oklo time, lying toward the near-end of the QSO range.  The argument raised, however, the question on how they chose this specific time $t_1$ out of the entire history of the meteorites.  This issue will be discussed in the next Section.

\section{Isochron analysis -- extended}

As we learned, most of the researchers working on the isochron analysis have taken it for granted that the decay rate $\lmd$ is a pure constant, having had none of the slightest idea on its possible time-dependence.  Notice, however, that the relation \reflef{isch-1}) is a consequence of the differential equation
\beq
\frac{dN_{\rm Re}(t)}{dt}= -\lmd(t)N_{\rm Re}(t),
\label{isch-7}
\eeq
where we have been immediately allowed to extend a constant $\lmd$ to a time-dependent $\lmd(t)$.  This equation permits an obvious solution
\beq
N_{\rm Re}(t)=A\exp\left[ -\int_{t_1}^t \lmd(t')dt' \right].
\label{isch-8}
\eeq
It is an easy exercise to substitute \reflef{isch-8}) into \reflef{isch-7}) finding the former solves the latter.

With $t=t_0$, we also put \reflef{isch-8}) into the form of \reflef{isch-1}),
\beq
N_{\rm Re}(t_0)=A e^{-\bar{\lmd}(t_0-t_1)},
\label{isch-9}
\eeq
where $\bar{\lmd}$ is defined by a time-average of $\lmd(t)$:
\beq
\bar{\lmd} =\frac{1}{t_0-t_1}\int_{t_1}^{t_0}\lmd(t')dt'.
\label{isch-10}
\eeq
In this way we have come to find that the decay rate which has been determined to the precision of 0.5\% before \cite{smol} is now re-interpreted as the time-varying decay rate averaged over the entire history of the meteorites \cite{yfai},
\beq
\Bigl| \frac{\Delta\bar{\lmd}}{\bar{\lmd}}  \Bigr| \lsim 5\times 10^{-3}.
\label{isch-5a}
\eeq
The notion of $\bar{\lmd}$ was discussed briefly in the final published version of Ref. \cite{olive2}.

\mbox{}\\[-5em]
\bfig[tbh]
\hspace*{8em}
\vspace{-9em}
\epsfxsize=10.4cm
\epsffile{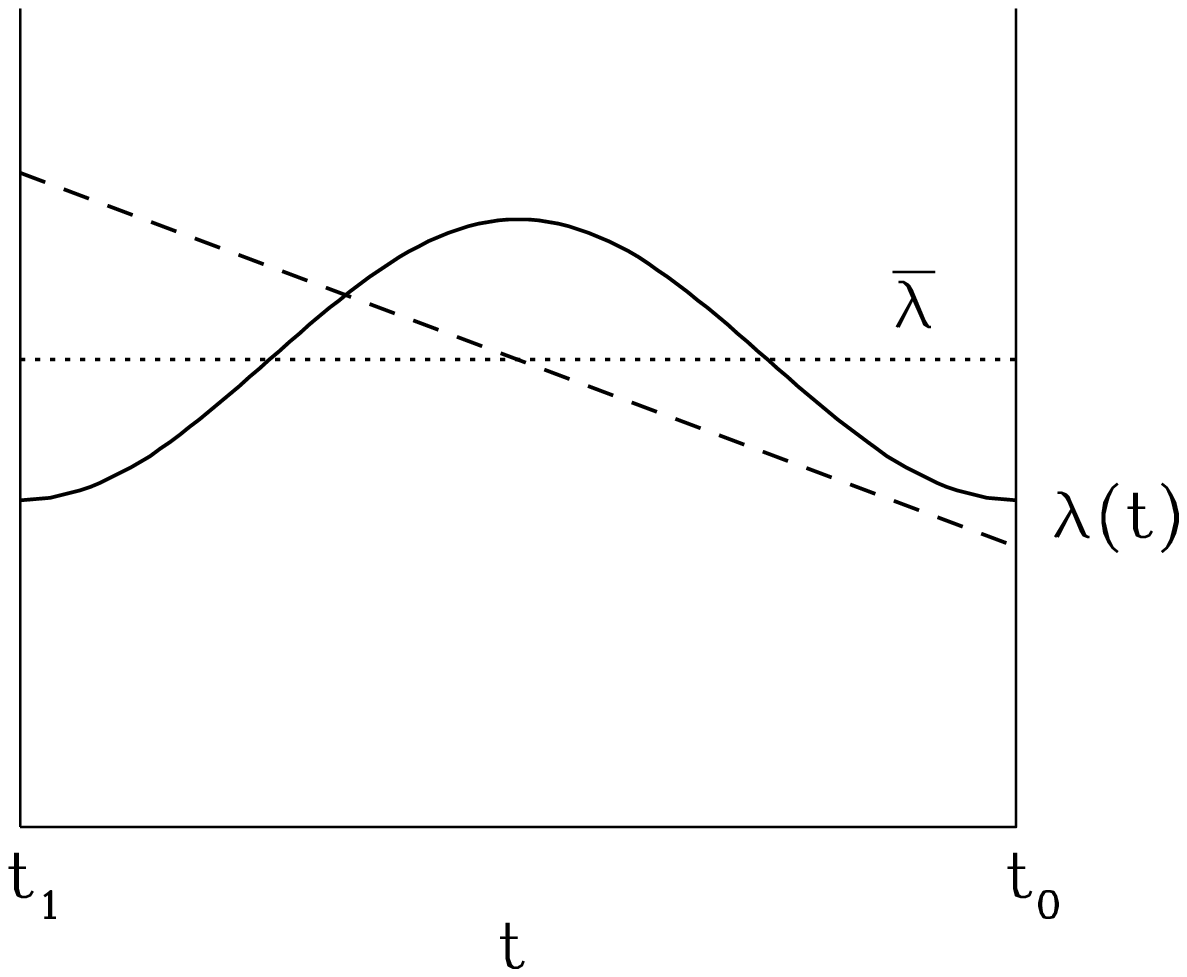}
\mbox{}\\[-4.9em]
\caption{Two random examples, a straight line and a sinusoidal curve, of the function $\lmd(t)$ for the same average $\bar{\lmd}$.}
\label{fig3}
\efig

For a given averaged $\bar{\lmd}$, we may think of many different ways of time-dependence of $\lmd(t)$, as illustrated in Fig. \ref{fig3}.  The average, no matter how precisely it may be measured, has nothing to do with how the function behaves.  It masks any variation during the entire history between $t_1$ and $t_0$.  In this sense, the dating technique made at the {\em present time} is of different nature from looking directly for the {\em past events}, like probing the ancient Oklo remnants or measuring the wavelength of the absorption lines coming from very distant objects as in the QSO observation.

Apart from the question in principle, however, two quantities, averaged integral and in-between behaviors may have some relationship in practice.  This ought to be model-dependent.  We intend to look into some details expecting to understand  better the physical implications.

Olive et al. noticed \cite{olive2} that today's observed value of $\lmd$ covers
$\bar{\lmd}$ with the uncertainty of about $(1.5-3.0)$\% \cite{todayslmd}.  They applied the 
simplest linear approximation, basically the same as the dashed straight line
 illustrated in Fig. \ref{fig3}, thus expecting $|\Delta\lmd/\lmd|\lsim
 (0.015-0.03)$ at the other end, $t \sim t_1$.  Aside from a fundamental
 question  on how the linear approximation is supported by any physical
 ground, we point out that the time-range during which $\lmd(t)$
 deviates from $\bar{\lmd}$ significantly is likely quite wide as we
 find in Fig. \ref{fig3}, in contrast either to a narrow time-range of
 the Oklo phenomenon or to nearly negligible horizontal error bars of
 each QSO data-point.  Even with certain elaboration suggested in
 \cite{olive2}, it seems by no means obvious how to confine the timing of the $\alpha$ variation into a narrow and well-defined range around the meteorite time, as also illustrated in Fig. 3 of \cite{pt} or in the Figure on page 61 of \cite{sciam} for the larger audience..

They also attempted to rely on the negative $\Delta\alpha/\alpha$
claimed by one of the QSO groups \cite{keck}.  We should be cautious, however, in view of the apparently contradicting reports by other groups \cite{qrl,indfr,lev}.  It is, as we observe, this type of conflict that is supposed to be solved by the meteorite constraint.  Also related to this issue, one of us (Y.F.) has even proposed to fit the QSO data in terms of an oscillating $\alpha (t)$,
which appears favored by the accelerating Universe as well as the Oklo
constraint accepted to be more solid than the meteorite constraint \cite{yf}-\cite{plb2}.  This may further contribute to weaken the constraint on the time of the $\alpha$ changes on the basis of the meteorite dating.

\bfig[htb]
\vspace{-2em}
\hspace*{10.5em}
\epsfxsize=7cm
\epsffile{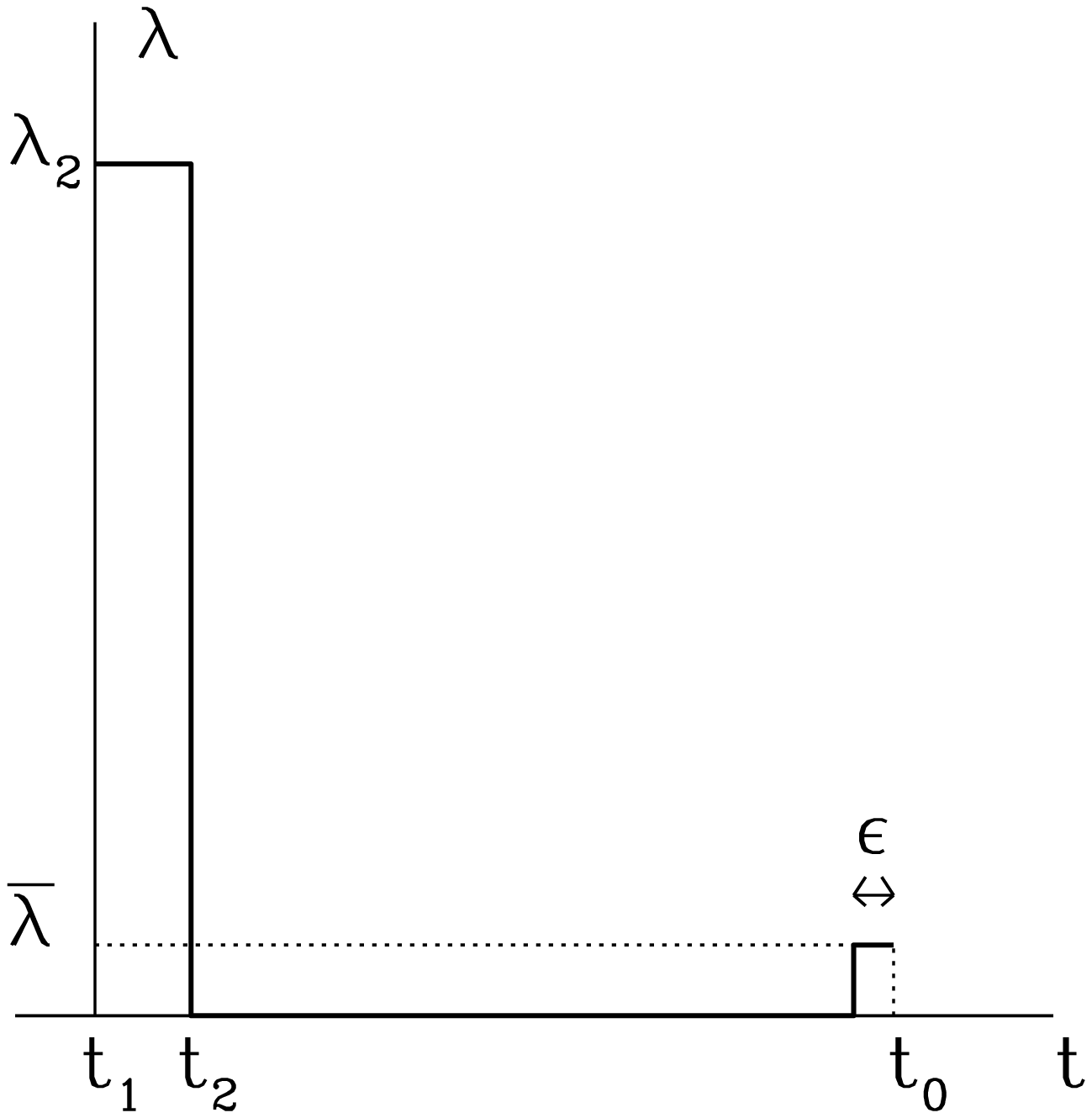}
\mbox{}\\[-5em]
\caption{ $\lmd_2$ is made the largest possible under the condition of the average $\bar{\lmd}$ and positivity of $\lmd(t)$.  $\epsilon$ can be chosen infinitesimally small. }
\label{tintv2}
\efig

\section{Consequences of the extended analysis}

We present examples to show how different consequences the preceding Section entails.

\subsection{Dyson's argument -- II}

We revisit Dyson's historical argument, which is expected to provide a lesson for other future studies.  Fig. \ref{tintv1} will be replaced by Fig. \ref{tintv2},  in which $\lmd (t)$ for $t>t_2$ has been set to zero, except for a small time-interval of the width $\epsilon$ toward the present time $t_0$.  This narrow portion is inserted technically to ensure that today's value is (close to) $\bar{\lmd}$, but $\epsilon$ can be chosen infinitesimal mathematically.  We chose $\lmd=0$
 between $t_2$ and $t_0-\epsilon$ in order for $\lmd_2$ to be as large as possible keeping the averaged value of the function $\lmd(t)$ to be $\bar{\lmd}$ under the natural condition $\lmd(t) \geq 0$.

Requiring the average to equal $\bar{\lmd}$ results in the condition, in the limit $\epsilon \rightarrow 0$:
\beq
(t_2 -t_1 )\lmd_2 =(t_0 -t_1) \bar{\lmd}.
\label{isch-15}
\eeq
The right-hand side, which can also be put into $(t_0 -t_1)/\bar{\lmd}^{-1}\approx 0.08 \ll 1$ , and so is the left-hand side, where we used $\bar{\lmd}^{-1}\approx \lmd_0^{-1}\approx 6\times 10^{10}{\rm y}$.  This implies that there was no time sufficiently long for rhenium to decay significantly before the time $t_2$, no matter how large $\lmd_2$, hence how small $\alpha$ might have been.  This illustrates how unlikely the naive argument, like the one due to Dyson, applies here, as far as the positivity condition of the decay rate is maintained.

This critical comment would not have followed if $\bar{\lmd}^{-1}$ were to happen unrealistically to be nearly comparable or shorter than $t_0-t_1$, the age of the meteorites, because then $(t_2-t_1)\lmd_2 \gsim 1$, and hence Re must have decayed out nearly entirely.  We also point out that the ``peak" between $t_1$ and $t_2$ can be anywhere inside the range $t_1$ and $t_0$.  This is one of the general features that knowing only the averaged decay rate tends to mask any details of the history.

\subsection{A simplified model relating $\Delta\bar{\lmd}/\bar{\lmd}$ to $\Delta\lmd/\lmd$}

In this subsection we try to apply the extended isochron formulation to find a theoretical model supposed to be realistic, still appealing to the simplified step-function approach.  We first point out that the straight line in the isochron plot, as in Fig. \ref{figisch1}, is an approximation after all.  In fact the isochron parameters, slope $S$ and the intercept $B$, for each group have been determined in such a way that they minimize the chi-squared for the data-points on the individual iron meteorites (samples) in the group, each with the error bar.  In Table 1, borrowed partly from Tables 1 and 2 of Ref. \cite{smol}, we show the number of samples and the estimated ages, relative to the absolute age of the angrites, as well as  the 2$\sigma$ errors for each group.\\[-2em]

\begin{table}[h]
\caption{Estimated ages and the number of samples of the groups \cite{smol}. }
\mbox{}\\
\hspace*{11em}
\begin{tabular}{c|cccc}
Group & IIA & IIIA & IVA &IVB \\   \hline
\rule{0mm}{4mm}Number of samples & 10 & 12 & 10 & 9 \\
Age \hspace{.7em}$(10^6{\rm y})$&4537 & 4558 & 4464 & 4525 \\
Error $(10^6{\rm y})$ & 8 & 12 & 26 & 29\\
\end{tabular}
\label{tab1}
\end{table}

These errors may be interpreted to be purely statistical in any conventional approach, but we dare to assume that individual iron meteorites in a group were indeed formed at different times distributed in the range indicated by the last row of Table \ref{tab1}, due to some physical processes.  This implies that a usual straight line in the isochron plot is only an approximate representation of 10 or so slightly different straight lines  for individual meteorites.

We also imagine that the different age yields different $\bar{\lmd}$ thanks to a specific way of the time-dependence of $\lmd$, as will be shown shortly.  We would then be able to relate $\Delta\bar{{\lmd}}/\bar{\lmd}$ to $\Delta\lmd /\lmd$.  Let us consider an individual meteorite belonging to a group $g$. Also assume a rectangular-type function for the time-dependent $\lmd(t)$:
\beq
\lmd(t)=\left\{
\begin{array}{ll}
\lmd_0+\delta =\lmd_2,\quad & \mbox{for}\quad t_3 \leq t \leq t_2, \nnb\\[.8em]
\lmd_0, &  \mbox{otherwise},
\end{array}
\right.
\label{isch-20-a}
\eeq
where $\delta$ can be positive or negative but subject to the condition $|\delta |\ll \lmd_0$.  We may have chosen different steps $\delta_2$ and $\delta_3$, respectively, at $t_2$ and $t_3$, but decided not to do so for simplicity.  Further complications might be avoided by choosing $t_2-t_3 \ll t_0-t_1$.  It then follows
\beq
\frac{\Delta\lmd}{\lmd}\approx \frac{\delta}{\lmd_0}.
\label{isch-21-a}
\eeq

Now we assume that one of the meteorites is formed at $t_{g1}$, which may differ from meteorite to meteorite but is confined to the range
\beq
t_g -w_g\lsim t_{g1} \lsim t_g+w_g,
\label{isch-22}
\eeq
where $t_g$ and $w_g$ are taken from the second and the third rows, respectively, of Table \ref{tab1}.  It might be convenient to call this range the ``initial $g$-range."   We also add another simplifying assumption  $t_2-t_3 \gsim 2w_g$.

\bfig[htb]
\vspace{-2em}
\hspace*{11em}
\epsfxsize=7cm
\epsffile{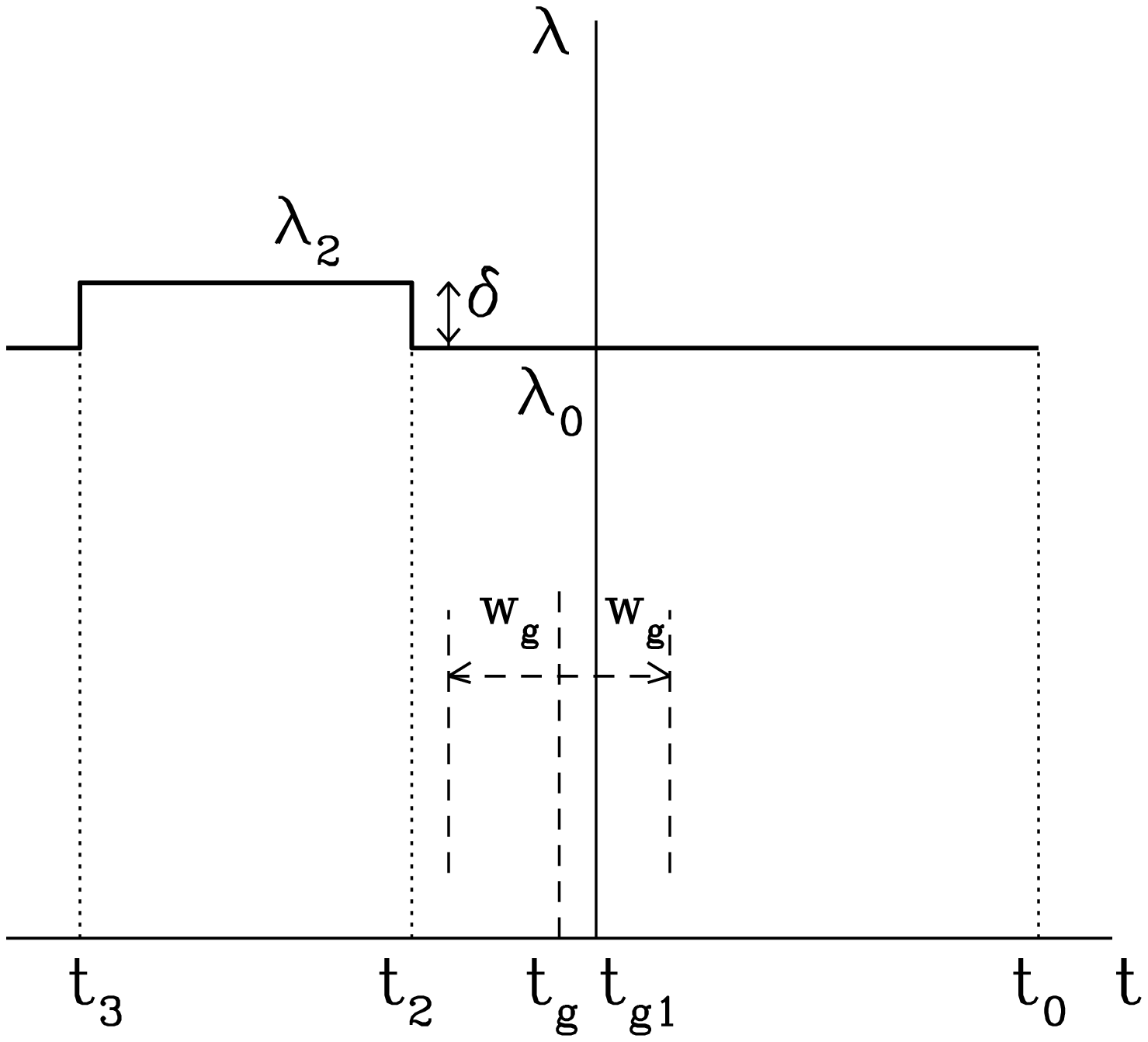}
\mbox{}\\[-5em]
\caption{ The change of $\alpha$ taking place between $t_2$ and $t_3$ before the initial $g$-range between $t_g-w_g$ and $t_g+w_g$.
 }
\label{tintv7}
\efig

Suppose first the change of $\lmd (t)$ had taken place before $t_g -w_g$, as illustrated in Fig. \ref{tintv7}.  Obviously the ``hump" (or dent if $\delta<0$) does not contribute to the integral which appears in \reflef{isch-8}) with $t=t_0$.  Suppose, however, the right-end $t_2$ of the hump enters the initial $g$-region between $t_g-w_g$ and $t_g+w_g$.  It may happen that $t_2 >t_{g1}$.  We thus expect a change of the integral:
\beq
\Biggl| \Delta \left( \int_{t_{g1}}^{t_0} \lmd(t') dt'\right)  \Biggl|= | (t_2-t_{g1})\delta |.
\label{isoch-25}
\eeq
Since the integral on the left-hand side defines $(t_0-t_{g1})\bar{\lmd}$, we find the change of $\bar{\lmd}$ as given by
\beq
|\Delta \bar{\lmd}|=\frac{(t_2-t_{g1})|\delta| }{t_0-t_{g1}},
\label{isoch-25-a}
\eeq
for a particular meteorite formed at $t_{g1}$.  This also results in a change of fluctuation of $\bar{\lmd}$ of the group arising from different ages of the individual meteorites also due to the time-dependence of $\lmd(t)$.  For an approximate estimate of the effect for the group $g$, we replace $t_2-t_{g1}$ by $w_g$ as far as $t_2$ happens to land in the initial $g$-range. Also by replacing $t_{g1}$ in the denominator by $t_g\approx t_1$, we obtain
\beq
\Biggl| \frac{\Delta \bar{\lmd}}{\bar{\lmd}} \Biggr|\approx \frac{w_g}{t_0-t_{1}}\frac{|\delta |}{\bar{\lmd}}\approx \frac{w_g}{t_0-t_{1}} \Biggl| \frac{\Delta \lmd}{\lmd} \Biggr|,
\label{isoch-25-b}
\eeq
where we have used \reflef{isch-21-a}) with the use of $\lmd_0 \approx \bar{\lmd}$.

Suppose then the hump of $\delta$ covers the entire initial $g$-range.  In addition to the integral between $t_g-w_g$ to $t_g+w_g$, we must also include the contribution coming from the range from $t_g+w_g$ to $t_2$.  The former contribution remains essentially the same as \reflef{isoch-25-b}) as far as the order of magnitude is concerned, while the latter is common to all the meteorites in the group $g$, hence fails to differentiate individual meteorites from each other.  We obtain basically the same result \reflef{isoch-25-b}).  In this sense the same contribution is obtained also when the left-end $t_3$ of the hump lands inside the initial $g$-range.  Finally no contribution appears to $\Delta\bar{\lmd}/\bar{\lmd}$ if the hump is entirely to the right of the initial $g$-range.

According to Table \ref{tab1}, we have $w_g \sim 8.0 \:\zeta \times 10^6\:{\rm y}$ where the coefficient $\zeta$ falls into the range roughly from $\sim 1$ for IIA to $\sim 3.6$ for IVB.  We thus estimate
\beq
\Bigl| \frac{w_g}{t_0-t_1} \Bigr| \approx 1.7 \:\zeta\times 10^{-3},\quad\mbox{with}\quad 1\lsim \zeta \lsim 3.6.
\label{isoch-28}
\eeq

We find it further necessary to estimate what portion of the ``observed" value 0.5\% comes from the $\delta$ effect.  We introduce another parameter $\eta <1$, in terms of which the left-hand side of \reflef{isoch-25-b}) is expressed by $5\eta\times 10^{-3}$ due to \reflef{isch-5a}).  We also substitute from \reflef{dys-3-1}) expressing $\Delta\lmd/\lmd$ in terms of $\Delta\alpha/\alpha$.  In this way we finally obtain
\beq
\Bigl| \frac{\Delta\alpha}{\alpha} \Bigr| \sim 1.7\times 10^{-4}\frac{\eta}{\zeta},
\label{isoch-29}
\eeq
which applies if $\alpha$ happens to be different from today's value in a range that {\em covers} the time of the meteorite formation, with a certain width $\sim w_g$.

Now with a tentative choice $\zeta =1$, we find $|\Delta\alpha/\alpha| \sim \eta\times 10^{-4}$.   If $\eta$ is close to one, the right-hand side of \reflef{isoch-29}) appears too large compared with the currently reported QSO data, certainly not  a promising result.  If choosing $\eta\approx 1$ were to turn out to yield somewhat smaller estimate than the  QSO result, we would have been ready to accept our model as providing a stronger constraint than the QSO.  Unfortunately that never happens to be the case.  We are then forced to assume $\eta$ much smaller than unity, implying, rather naturally, that a major part of the uncertainty comes from measurement errors, for example.  But then the result will be inevitably less useful.

According to our previously repeated criticism, the analysis in terms of the time-averaged decay rate fails generally to locate precisely when $\alpha(t)$ changed.  But  a specific way of constructing the model has allowed us to focus selectively upon a narrow range around the meteorite time for $\alpha (t)$ which can have changed over a much wider range, though the resulting $\Delta\alpha/\alpha$ is constrained in a way quite different from many of the related references.  We will also comment in the next Section briefly on using \reflef{isoch-29}) as a test of overall consistency of the analysis.

\section{Relevance to the Oklo and the QSO}

As we showed in Section 1, studying the meteorite constraint was motivated to give a solution to an apparent conflict between the Oklo and the QSO.  From this point of view it seems pertinent to explain briefly what our overall scenario can be like, with special attention to \reflef{isoch-29}).

Notice first that, as was emphasized in Section 1, the ``error-bar" for the Oklo constraint is at the level of $10^{-3}-10^{-2}$ in units of $y=(\Delta\alpha/\alpha)\times 10^5$, as used in Fig. \ref{summ}.  If we extrapolate the horizontal straight line $y=-0.573$ for Keck/HIRES \cite{keck} down to $s=0.142$, this single point contributes $\chi^2$ more than hundreds, too large for the entire fit including the 143 QSO data-points to be acceptable. It was argued that the scalar field which is responsible for the time-dependent $\alpha$ at Oklo on the Earth is different from the one for the far distant objects \cite{barrow}.   We find this, however, unlikely to be the case partly because we expect the space-dependent component of the scalar field to show the nature of finite force-range \cite{yfptp} in connection with non-Newtonian gravity. Rather we prefer a curved fitting function.

A reasonable assumption might be provided by a damped oscillation \cite{mzn},
\beq
y(s)= a\left(  e^{b(s-1)} \cos(v-v_1)-e^{-b} \cos(v_1) \right),
\label{lev-1}
\eeq
where $v/s = v_{\rm oklo}/s_{\rm oklo}=2\pi T^{-1}$ with the coefficient $v_1$ determined by
\beq
v_1 = \tan^{-1}\left( \left( e^{-bs_{\rm oklo}} -\cos(v_{\rm oklo})\right) /\sin(v_{\rm oklo}) \right),
\label{lev-2}
\eeq
due to which $y(s)$ vanishes naturally at $s=0$ (today) and $s=s_{\rm oklo}\approx 0.142$, as a good approximation to a very small value.  We admit that this near identification of the Oklo with a zero of oscillation is provisional, still believing that the oscillation is supported by the accelerating Universe.

Accepting \reflef{lev-1}) as a phenomenological function, at this moment, the best fit for the data set in \cite{keck} is obtained \cite{mzn} for $a= 0.020, b= 5.5$, and $T=1.352$ with $\chi_{\rm rd}^2=1.015$, to be compared with $\chi_{\rm rd}^2=1.023$ for the original weighted-mean fit.  Then at $s=0.33$, the meteorite time, we find $y= -0.072$  which is about 3 times as large as the meteorite constraint $|y|= 0.025$ corresponding to \reflef{isch-6}).  In view of the non-uniqueness of this constraint as discussed before, no further effort appears to be worth attempting for $s$ between the Oklo and the QSO times.  We also find that trying to lessen the calculated $|y|$ for this range of $s$ proves increasingly difficult unless it departs from the observed flat behavior excessively toward the high-$s$ end \cite{wagg,yfai}.

The same technique applied to the data set of \cite{indfr} reveals a different feature.  The original weighted-mean analysis gave nearly a null result, whereas  the data is best fit \cite{mzn} by the nonzero function \reflef{lev-1}) with $a=-0.050, b=3.1$ and $T=0.134$ yielding $\chi_{\rm rd}^2=0.53$, even better than the value 0.95 in \cite{indfr}.  The oscillatory behavior, being quite different from the nearly flat distribution in the data of \cite{keck}, is already apparent but can be easily washed out by a 1-parameter fit in terms of a horizontal straight line.  Our fit gives $y=0.166$ at $s=0.33$, reaching close to a peak of the oscillation, however, with nearby zeros at $s=0.28$ and $s=0.36$.  The meteorite constraint seems unlikely to play any crucial role even in its less restricted form.

Having estimated $\Delta\alpha/\alpha$ from the QSO fits, we may substitute the results back to the left-hand side of \reflef{isoch-29}) to evaluate $\eta$, finding $(0.5-1.8)\times 10^{-2}$ and $(1.2-4.3)\times 10^{-2}$ for \cite{keck} and \cite{indfr}, respectively.  No unique conclusion can be likely drawn from these numbers.  We still hope that future more detailed analysis of the same kind might play a decisive role in selecting the true time-dependence of $\alpha$ obtained from the QSO absorption lines.

We finally add how \reflef{lev-1}) is related to the cosmological acceleration.   We start with pointing out that the scalar-tensor theory provides a successful implementation of the scenario of a decaying cosmological constant \cite{cup}, according to which the observed $\Lmd(t)$, time-dependent cosmological ``constant" in the Einstein conformal frame, behaves like $\sim t^{-2}$ (in the reduced Planckian unit system), allowing us to understand the smallness by 120 orders of magnitude without appealing to an extreme fine-tuning of parameters.  Notice $t_0 \sim 10^{60}$ in units of the Planckian time. In order further to understand the nearly constant behavior of $\Lmd$ rather than falling off like $t^{-2}$, we need another mechanism to keep the scalar field to stay nearly standstill.  Probably the easiest way is to assume that the scalar field is trapped by a potential which naturally causes a damped oscillation to be identified with what we find to fit the QSO data.  This view is consistent, as we interpret \cite{plb2},  with the result of a careful analysis in Refs. \cite{qrl,lev}, which also appears to be consistent with the behavior encountered in \cite{indfr}.

Now from a more theoretical point of view keeping a strong tie with string theory than the phenomenology-oriented approach of quintessence \cite{quint}, the scalar-tensor theory may not allow us to ignore the emergence of a local field responsible for non-Newtonian gravity which features an intermediate force-range of macroscopic size and violation of Weak Equivalence Principle (WEP). This is also expected to free us from the long-standing constraint from the solar-system experiment. For more details on these crucial issues, see Refs. \cite{cup,bls}.

It should be truly remarkable and fascinating if these different kinds of phenomena are related to possible time-variability of the fine-structure constant via a scalar field.  We want to see if the issue of the rhenium decay contributes to verify these interrelations. \\[.5em]

\noindent
{\Large\bf Acknowledgments}
\mbox{}\\

We thank Hisayoshi Yurimoto for his invaluable help to learn details on meteorites, and Hiroshi Hidaka for his comment on the post-reactor contamination in the Oklo remnant.  Our thanks are also due to Yasushi Takahashi for his reading part of the manuscript.

\end{document}